# Two step disordering of the vortex lattice across the peak effect in a 3-dimensional type II superconductor $Co_{0.0075}NbSe_2$

Somesh Chandra Ganguli[a], Harkirat Singh[a], Garima Saraswat[a], Rini Ganguly[a], Vivas Bagwe[a], Parasharam Shirage[b], Arumugam Thamizhavel[a] and Pratap Raychaudhuri[a*]

[a] *Tata Institute of Fundamental Research, Homi Bhabha Road, Colaba, Mumbai 400005, India.*

[b] *Indian Institute of Technology Indore, IET-DAVV Campus, Khandwa Road, Indore 452017, India.*

The vortex lattice in a Type II superconductor provides a versatile model system to investigate the order-disorder transition in a periodic medium in the presence of random pinning. Here, using scanning tunnelling spectroscopy in a weakly pinned $Co_{0.0075}NbSe_2$ single crystal, we show the possibility of a two-step order-disorder transition of the vortex lattice in a 3-dimensional superconductor at low temperatures across the magnetic field driven peak effect. At the onset of the peak effect, the equilibrium quasi-long range ordered state transforms into an orientational glass through the proliferation of dislocations. At a higher field, the dislocations dissociate into isolated disclination giving rise to an amorphous vortex glass. We also show the existence of a variety of additional non-equilibrium metastable states, which can be accessed through different thermomagnetic cycling.

[*] E-mail: pratap@tifr.res.in

Understanding the evolution of the structure of the vortex lattice (VL) in a weakly pinned type II superconductor is of paramount importance since it determines superconducting properties that are directly relevant for applications, i.e. critical current and the onset of electrical resistance. Over the past two decades, there have been intense efforts to understand the nature of the order-disorder transition of the VL with temperature or magnetic field[1,2,3]. It is generally accepted that in a clean system the hexagonal VL realised at low temperature and magnetic field, can transform to vortex liquid above a characteristic temperature ($T$) and magnetic field ($H$). Random pinning, arising from crystalline imperfection in the superconductor significantly complicates this scenario. It has been argued that since the system can no longer sustain true long-range order, both the ordered and the disordered state can become of glassy nature[4,5], characterised by different degree of positional and orientational order. In addition, the VL can exist in a variety of non-equilibrium metastable states[6,7], depending on the thermomagnetic history of the sample.

In contrast to the VL in superconducting thin films, where the order-disorder transition can be understood within the framework of Berezinski-Kosterlitz-Thouless (BKT) theory of 2-dimensional (2-D) melting[8,9,10], the VL in a 3-dimensional (3-D) superconductor presents a more challenging problem. In this case the vortex line is rigid only up to a length scale much shorter than sample dimensions. Thus in a weakly pinned single crystal, the vortex line can bend considerably along the length of the vortex. It is generally accepted that in the presence of weak pinning the Abrikosov VL can transform into a quasi long range ordered state such as Bragg glass[11] (BG), which retains long-range orientational order of a perfect hexagonal lattice but where the positional order decays algebraically with distance. Theoretically, both the possibility of a direct first order transition from a BG to a vortex glass (VG) state[12,13] (with short range positional and orientational order) as well as transitions through an intermediate state, such as multi-domain glass or a hexatic glass[14,15], have been discussed in the literature. While many experiments find evidence of a first-order order-disorder transition[16,17,18,19], additional continuous transitions and crossovers have been reported in other regions[20,21,22] of the *H*-*T* parameter space, both in low-$T_c$ conventional superconductors and in layered high-$T_c$ cuprates.

Experimentally, the order-disorder transition in 3-D superconductors has been extensively studied through bulk measurements, such as critical current[23], ac susceptibility[24,25] and dc magnetisation[26,27]. These studies rely on the fact that in the presence of random pinning centres, the VL gets more strongly pinned to the crystal lattice as the perfect hexagonal order of the VL is relaxed[28]. The order-disorder transition thus manifests as sudden non-monotonic enhancement of bulk pinning[29], and consequently of the critical current and the diamagnetic response in ac susceptibility measurements. Known as the "peak effect", this has been a central theme of many studies on the static and dynamic properties of VLs. These measurements, although valuable in establishing the phase diagram of type II superconductors, do not reveal the evolution of the microscopic structure of the VL across the order-disorder transition. A more direct, though less used method, is through direct imaging of the VL using scanning tunnelling spectroscopy[30,31,32,33,34] (STS). The main challenge in this technique is to get large area images that are representative of the VL in the bulk crystal.

Here, we track the evolution of the equilibrium state of the VL across the magnetic field driven peak effect at low temperature using direct imaging of the VL using STS in an $NbSe_2$ single crystal, intercalated with 0.75% of Co. The intercalated Co atoms act as random pinning centres, making the peak effect more pronounced compared to pure $NbSe_2$ single crystal[35]. We analyse VL images consisting of several hundred to a thousand vortices at 350 mK, taken across the magnetic field driven the peak effect. For low fields the stable state of the VL has nearly perfect hexagonal structure, with long range orientational order and a slowly decaying positional order. At the onset of the peak effect, dislocations proliferate in the VL, transforming the VL to an orientational glass (OG) with slowly decaying orientational order. Above the peak of the peak effect, dislocations dissociate into isolated disclinations driving the VL into an amorphous vortex glass (VG) which connects smoothly to the liquid state close to the upper critical field, $H_{c2}$. Our results points towards the possibility that the vortex lattice disorders in two steps, through successive destruction of positional and orientational order.

**Results**

**Bulk pinning properties.** Fig.1(a) shows the bulk pinning response of the VL at 350 mK, measured from the real part of the linear ac susceptibility ($\chi'$) (see Supplementary Information) when the sample is cycled through different thermomagnetic histories. The $\chi'$-$H$ for the zero field cooled (ZFC) state (red line) is obtained while ramping up the magnetic field after cooling the sample to 350 mK in zero magnetic field. The “peak effect” manifests as a sudden increase in the diamagnetic response between 16 kOe ($H_p^{on}$) to 25 kOe ($H_p$) after which $\chi'$ monotonically increases up to $H_{c2}$ ~ 38 kOe. When the magnetic field is ramped down after reaching a value $H > H_{c2}$ (black line, henceforth referred as the ramp down branch), we observe a hysteresis starting below $H_p$ and extending well below $H_p^{on}$. A much more disordered state of the VL with stronger diamagnetic response is obtained when the sample is field cooled (FC), by applying a field at 7 K and cooling the sample to 350 mK in the presence of the field (solid squares). This is however a non-equilibrium state: When the magnetic field is ramped up or ramped down from the pristine FC state, $\chi'$ merges with the ZFC branch or the ramp down branch respectively. In contrast, $\chi'$ for the ZFC state is reversible with magnetic field cycling up to $H_p^{on}$, suggesting that it is the more stable state of the system. Fig. 1(b) shows the phase diagram with $H_p^{on}$, $H_p$ and $H_{c2}$, obtained from isothermal χ'-$H$ scans at different temperatures.

**Real space imaging of the VL.** The VL is imaged using STS over a 1 μm × 1 μm area close to the center of the cleaved crystal surface. We first focus on the VL along the ZFC branch. Figure 2 shows the representative conductance maps over the full scan area at 15 kOe and 24 kOe where vortices manifest as a local minima in the conductance. Figure 3 (a)-(f) show the conductance maps superposed with the Delaunay triangulated VL for 6 representative fields, when the magnetic field is ramped up at 350 mK in the ZFC state. The Fourier transforms (FT) corresponding to the unfiltered images are also shown. We identify 3 distinct regimes. For $H < H_p^{on}$, the VL is free from topological defects and the FT show 6 bright spots characteristic of a hexagonal lattice. Between $H_p^{on} < H \leq H_p$ dislocations (pairs of nearest neighbor lattice points with 5-fold and 7-fold coordination) gradually proliferate in the system. We call this state an orientational glass (OG). The dislocations do not completely destroy the orientational order which can be seen from FTs which continue to display a six-fold symmetry. For $H > H_p$ the disclinations (isolated lattice points with 5-fold or 7-seven fold coordination) proliferate in the system driving the VL into an isotropic VG. The FT shows a ring, characteristic of an isotropic disordered state. We observe a significant range of phase coexistence[36], where both large patches with dislocations coexist with isolated disclinations. Going to higher fields, 32 and 34 kOe (Fig. 3 (g)-(h)) we observe that the VL gets gradually blurred to form a randomly oriented linear structures, where the vortex lines start moving along preferred directions determined by the local surrounding. This indicates a softening of the VG and a gradual evolution towards the liquid state close to $H_{c2}$.

Further quantitative information on this sequence of disordering is obtained from the orientational and positional correlation functions, $G_6(\bar{r})$ and $G_{\bar{K}}(\bar{r})$, which measure the degree of misalignment of the lattice vectors and the relative displacement between two vortices separated by distance $r$ respectively, with respect to the lattice vectors of an ideal hexagonal lattice. The orientational correlation function is defined as, $G_6(r) = (1/n(r,\Delta r))\left(\sum_{i,j}\Theta\left(\frac{\Delta r}{2} - \left|r - \left|\bar{r}_i - \bar{r}_j\right|\right|\right)\cos 6\left(\theta(\bar{r}_i) - \theta(\bar{r}_j)\right)\right)$, where $\Theta(r)$ is the Heaviside step function, $\theta(\bar{r}_i) - \theta(\bar{r}_j)$ is the angle between the bonds located at $\bar{r}_i$ and the bond located at $\bar{r}_j$, $n(r,\Delta r) = \sum_{i,j}\Theta\left(\frac{\Delta r}{2} - \left|r - \left|\bar{r}_i - \bar{r}_j\right|\right|\right)$, $\Delta r$ defines a small window of the size of the pixel around $r$ and the sums run over all the bonds. We define the position of each bond as the coordinate of the mid-point of the bond. Similarly, the spatial correlation function, $G_{\bar{K}}(r) = (1/N(r,\Delta r))\left(\sum_{i,j}\Theta\left(\frac{\Delta r}{2} - \left|r - \left|\bar{R}_i - \bar{R}_j\right|\right|\right)\cos \bar{K}\cdot\left(\bar{R}_i - \bar{R}_j\right)\right)$, where ***K*** is the reciprocal lattice vector obtained from the Fourier transform, $R_i$ is the position of the *i*-th vortex, $N(r,\Delta r) = \sum_{i,j}\Theta\left(\frac{\Delta r}{2} - \left|r - \left|\bar{R}_i - \bar{R}_j\right|\right|\right)$ and the sum runs over all lattice points. We restrict the range of $r$ to half the lateral size (1 μm) of each image, which corresponds to $11a_0$ (where $a_0$ is the average lattice constant) at 10 kOe and $17a_0$ for 30 kOe. For an ideal hexagonal lattice, $G_6(r)$ and $G_{\bar{K}}(r)$ shows sharp peaks with unity amplitude around 1$^{st}$, 2$^{nd}$, 3$^{rd}$ etc… nearest neighbour distance for the bonds and the lattice points respectively. As the lattice disorder increases, the amplitude of the peaks decay with distance and neighbouring peaks at large $r$ merge with each other.

Figure 4(a) and 4(b) show the $G_{\bar{K}}(r)$ (averaged over the 3 principal ***K*** directions) and $G_6(\bar{r})$, calculated from individual VL images, as a function of $r/a_0$ for different fields. At 10 kOe and 15 kOe, $G_6(r)$ saturates to a constant value of ~0.93 and ~0.86 respectively after 2-3 lattice constants, indicating that the presence of long-range orientational order. The envelope of $G_{\bar{K}}(r)$ decays slowly but almost linearly with *r*. Since the linear decay cannot continue for large *r*, this reflects our inability to capture the asymptotic behaviour at large *r* at low fields due to limited field of view. While we cannot ascertain whether $G_{\bar{K}}(r)$ decays as a power-law for large *r* as predicted for a BG, the slow decay of $G_{\bar{K}}(r)$ combined with the long-range orientational order is indicative of quasi long-range positional order (QLRPO). In the OG state (20-25 kOe), $G_6(r)$ decays slowly with increasing *r*. The decay of $G_6(r)$ with *r* is consistent with a power-law ($G_6(r) \propto 1/r^{\eta}$), characteristic of quasi-long-range orientational order (Fig. 4(c)). $G_{\bar{K}}(r)$, on the other hand displays a more complex behaviour. At 20 kOe, within our field of view the $G_{\bar{K}}(r)$ envelope decays exponentially with positional decay length, $\xi_p$ ~ 6.7. However for 24 and 25 kOe where the initial decay is faster, we observe that the exponential decay is actually restricted to small values of $r/a_0$ (Fig. 4(d)), whereas at higher values $G_{\bar{K}}(r)$ decays as a power-law (Fig. 4(e)). The OG state thus differs from the QLRPO state in that it does not have a true long-range orientational order. It also differs from the hexatic state in 2-D systems, where $G_{\bar{K}}(r)$ is expected to decay exponentially at large distance. Similar variation of $G_{\bar{K}}(r)$ has earlier been reported at intermediate fields in the VL of a neutron irradiated $NbSe_2$ single crystal[32], although the data in that case did not extend to the VG state. Finally, above 26 kOe, $G_6(r) \propto e^{-r/\xi_{or}}$ ($\xi_{or}$ is the decay length of orientational order), giving rise to regular amorphous VG state with short-range positional and orientational order (Fig. 4(f)).

The possibility of a state with hexatic correlations between the onset and the peak of the "peak effect" has earlier been suggested from the field variation of the positional correlation length of the VL parallel ($\xi^{\parallel}$) and perpendicular ($\xi^{\perp}$) to the reciprocal lattice vector ***K***, from neutron scattering studies in Nb single crystal[37]. In that measurement, $\xi^{\parallel}$ and $\xi^{\perp}$ was inferred from the radial width and azimuthal width of the six first order scattering in the Ewald sphere, projected on the plane of the detector[38]. While the relation between these correlation length and the ones obtained from the decay of $G_{\overline{K}}(r)$ or $G_6(r)$ is not straightforward, it is nevertheless instructive to compare the corresponding correlation lengths obtained from our data. We obtain the corresponding lengths in our experiment from the average radial ($\Delta k_{||}$) and azimuthal ($\Delta k_{\perp}$) width of the six first-order Bragg peaks of the reciprocal lattice of each VL, obtained by taking the Fourier transform of a binary map constructed using the position of the vortices (See Methods and Supplementary material). Since the Bragg peaks fit well to a Lorentzian profile, we correct for the peak broadening arising from the finite size of the images and the position uncertainly arising from the finite pixel size of the images by subtracting the peak width of the Bragg spots of an ideal hexagonal lattice of the same size constructed on a grid containing the same number of pixels. Fig. 4 (g)-(h) show the variation of $\xi^{||} \sim 1/\Delta k_{||}$ and $\xi^{\perp} \sim 1/\Delta k_{\perp}$ with magnetic field. We observe that $\xi^{\perp}/a_0$ and $\xi^{||}/a_0$ decreases rapidly between $H_p^{on}$ and $H_p$ signalling the progressive decrease in orientational and positional order. At $H_p$, both $\xi^{\perp}/a_0$, $\xi^{||}/a_0 \sim 1$, showing both positional and orientational order are completely lost. Both the magnitude and field dependence of $\xi^{\perp}$ and $\xi^{||}$ are qualitatively consistent with ref. 37 even though the measurements here are performed at a much lower temperature.

We now discuss the ramp down branch focussing on the hysteresis region in the $\chi'$–$H$ measurements. Fig. 5(a)-(c) shows the VL configurations for the ramp down branch at 25, 20 and 15 kOe. The VL structures for the ramp down branch are similar to ZFC: At 25 and 20 kOe the VL shows the presence of dislocations and at 15 kOe it is topologically ordered. In Fig. 4(d)-(f) we compare $G_6(r)$ and $G_{\overline{K}}(r)$ calculated for the ZFC and the ramp down branch. At 25 kOe $\approx H_p$, we observe that $G_{\overline{K}}(r)$ for ZFC and ramp down branch are similar whereas $G_6(r)$ decays faster for the ramp down branch. However, analysis of the data shows that in both cases $G_6(r)$ decays as a power-law (Fig. 5(f)) characteristic of the OG state. At 15 kOe, which is just below $H_p^{on}$, both ZFC and ramp down branch show long-range orientational order, while $G_{\bar{K}}(r)$ decays marginally faster for the ramp down branch. Thus, while the VL in the ramp down branch is more disordered, our data do not provide any evidence of supercooling across either LQRPO→OG or OG→VG transitions as expected for a first order phase transition. Therefore, we attribute the hysteresis to the inability of the VL to fully relax below the transition in the ramp down branch.

We can now follow the magnetic field evolution of the FC state (Fig. 6). The FC state show an OG at 10 kOe (not shown) and 15 kOe (free dislocations), and a VG above 20 kOe (free disclinations). The FC OG state is however extremely unstable. This is readily seen by applying a small magnetic pulse (by ramping up the field by a small amount and ramping back), which annihilates the dislocations in the FC OG (Fig. 7) eventually causing a dynamic transition to the QLRPO state. It is interesting to note that metastability of the VL persists above $H_p$ where the ZFC state is a VG. The FC state is more disordered with a faster decay in $G_6(r)$ (Fig. 6, lower panels), and consequently is more strongly pinned than the ZFC state.

**Discussion**

The two step disordering observed in our experiment is reminiscent of the two-step melting observed in 2-D systems[39], where a hexatic fluid exists as an intermediate state between the solid and the isotropic liquid. However, the situation for the weakly pinned 3-D VL is more complex. Here, the reduced influence of thermal fluctuations prevents from establishing a fluid state in the presence of a pinning potential. Thus, the OG and the VG state are both glassy states with either frozen or very slow kinetics till very close to $H_{c2}$. The glassy nature of the VL also manifests by producing a number of non-equilibrium states, such as the OG below $H_p^{on}$ and the VG state below $H_p$ when the sample is field cooled.

One pertinent question in this context is the nature of the order-disorder transition across the peak effect. Within our definition, OG state is distinguished from the QLRPO state through the proliferation of dislocations which destroys the true long-range orientational order. While the role of topological defects on the QLRPO is at present is not completely understood. In the case of a BG it has been argued that a small number of dislocations need not necessarily destroy the BG state as long as the average distance between dislocations is much larger than the effective correlation length of the BG state[4]. If that is the case, then the transition to an OG might happen at a field higher than $H_p^{on}$ when the average separation between dislocations becomes of the order of $\xi^{\parallel}$, $\xi^{\perp}$. This issue is at present unresolved and needs to be settled through imaging of much larger lattices. On the other hand, the boundary between OG to VG seems to be well demarcated by a change in the decay in $G_6(r)$ from power-law to exponential. However, it is interesting to note that we do not observe any evidence of supercooling across the peak effect as expected for a first-order phase transition. Theoretically the possibility of a second order field induced melting transition has been speculated[40], though detailed calculations do not exist. Finally, the gradual softening of the VG as the magnetic field approaches $H_{c2}$ supports the view[13] that the VG and the vortex liquid are thermodynamically identical states in two different limits of viscosity.

The two-step disordering observed in our experiment bears some similarity with the phase diagram proposed in ref. 14 where the BG disorders to a VG through an intermediate "multidomain glass" (MG) state. However, in a MG the spatial correlations are similar to a BG up to a characteristic length-scale followed by a more rapid decay at larger distance. In contrast, in the intermediate OG state observed in our experiment, $G_{\overline{K}}(r)$ decays rapidly at short distance followed by a more gradual behaviour. The possibility of the BG transforming to an intermediate state that retains hexatic correlations, which would be closer to our experimental findings, has also been proposed as a possibility[14,40], but has not been explored in detail.

In summary, the physical picture emerging from our measurements is that across the peak effect, the VL in a weakly pinned Type II superconductor can disorder in two steps: From a QLRPO state at low field to an OG, and then from the OG to an isotropic VG which smoothly connects to the liquid state. Our experiments do not provide evidence of a first order order-disorder transition across the field driven peak effect at low temperatures. In this context we would like to note that a hexatic VL state (similar to the OG state), believed to represent the "ordered state" of vortex matter, has earlier been observed from magnetic decoration experiments on a layered high-$T_c$ $Bi_{2.1}Sr_{1.9}Ca_{0.9}Cu_2O_{8+\delta}$ single crystals at low temperatures and very low fields[41]. It would therefore be interesting to carry out further experiments on other low-$T_c$ and high-$T_c$ superconductor with different levels disorder which would provide valuable insight on the range of parameter space over which the OG state is observed in systems with different disorder and anisotropies.

## Material and Methods

**Sample preparation.** The $Co_{0.0075}NbSe_2$ single crystal was grown by iodine vapour transport method starting with stoichiometric amounts of pure Nb, Se and Co, together with iodine as the transport agent. Stoichiometric amounts of pure Nb, Se and Co, together with iodine as the transport agent were mixed and placed in one end of a quartz tube, which was then evacuated and sealed. The sealed quartz tube was heated up in a two zone furnace for 5 days, with the charge-zone and growth-zone temperatures kept at, $800^0$ C and $720^0$ C respectively. We obtained single crystals of nominal composition $Co_{0.0075}NbSe_2$ with lateral size of 4-5 mm. The crystal on which the measurements were performed had a superconducting transition temperature, $T_c$ ~ 5.3 K with a resistive transition width of 200 mK (see Supplementary material) and $H_{c2}$ ~ 38 kOe (for H ⊥ Nb planes). The same sample was used for susceptibility and STS measurements.

**a.c. suspectibility measurements.** a.c. susceptibility measurements were performed down to 350 mK using a home-built susceptometer, in a $^3$He cryostat fitted with a superconducting solenoid. Both ac and dc magnetic field were applied perpendicular to the Nb planes. The dc magnetic field was varied between 0-60 kOe. The ac excitation field was kept at 10 mOe where the susceptibility is in the linear response regime (see Supplementary material).

**Scanning tunneling spectroscopy measurements.** The VL was imaged using a home-built scanning tunneling microscope[42] (STM) operating down to 350 mK and fitted with an axial 90 kOe superconducting solenoid. Prior to STM measurements, the crystal is cleaved in-situ in vacuum, giving atomically smooth facets larger than 1 μm × 1 μm. Well resolved images of the VL are obtained by measuring the tunneling conductance ($G(V) = dI/dV$) over the surface at a fixed bias voltage ($V$~ 1.2 mV) close to the superconducting energy gap, such that each vortex core manifests as a local minimum in $G(V)$. Each image was acquired over 75 minutes after waiting for 15 minutes after stabilizing to the magnetic field. The precise position of the vortices are obtained from the images after digitally removing scan lines (see supplementary material) and finding the local minima in $G(V)$ using WSxM software[43]. To identify topological defects, we Delaunay triangulated the VL and determined the nearest neighbor coordination for each flux lines. Topological defects in the hexagonal lattice manifest as points with 5-fold or 7-fold coordination number. Since, the Delaunay triangulation procedure gives some spurious bonds at the edge of the image we ignore the edge bonds while calculating the average lattice constants and identifying the topological defects. Unless otherwise mentioned VL images are taken over an area of 1 μm × 1 μm. The correlation functions $G_6$ and $G_K$ are calculated using individual images.

---

[1]Brandt, E. H. The flux-line lattice in superconductors. *Rep. Prog. Phys.* **58,** 1465-1594 (1995).

[2]Higgins, M. J. & Bhattacharya, S. Varieties of dynamics in a disordered flux-line lattice. *Physica C* **257,** 232-254 (1996).

[3] Paltiel, Y. *et al.* Dynamic instabilities and memory effects in vortex matter. *Nature* **403,** 398-401 (2000).

[4]Giamarchi, T. & Le Doussal, P. Elastic theory of flux lattices in the presence of weak disorder. *Phys. Rev. B* **52,** 1242-1270 (1995).

[5] Fisher, D. S., Fisher, M. P. A. & Huse, D. A. Thermal fluctuations, quenched disorder, phase transitions, and transport in type-II superconductors. *Phys. Rev. B* **43,** 130-159 (1991).

[6] W. Henderson, E. Y. Andrei, M. J. Higgins and S. Bhattacharya, Metastability and Glassy Behavior of a Driven Flux-Line Lattice. *Phys. Rev. Lett.* **77,** 2077-2080 (1996).

[7] Pasquini, G., Pérez Daroca, D., Chiliotte, C., Lozano, G. S. & Bekeris, V. Ordered, Disordered, and Coexistent Stable Vortex Lattices in $NbSe_2$ Single Crystals. *Phys. Rev. Lett.* **100,** 247003 (2008).

[8] Berghuis, P., van der Slot, A. L. F. & Kes, P. H. Dislocation-mediated vortex-lattice melting in thin films of a-$Nb_3Ge$. *Phys. Rev. Lett.* **65,** 2583-2586 (1990).

[9] Yazdani, A. *et al.* Observation of Kosterlitz–Thouless-type melting of the disordered vortex lattice in thin films of alpha-MoGe. *Phys. Rev. Lett.* **70,** 505-508 (1993).

[10] Guillamón, I. *et al.* Direct observation of melting in a two-dimensional superconducting vortex lattice. *Nat. Phys.* **5,** 651-655 (2009).

[11] Klein, T. *et al.* A Bragg glass phase in the vortex lattice of a type II superconductor. *Nature* **413,** 404-406 (2001).

[12] Menon, G. I. & Dasgupta, C. Effects of pinning disorder on the correlations and freezing of the flux liquid in layered superconductors. *Phys. Rev. Lett.* **73**, 1023-1026 (1994).

[13] Kierfeld, J. & Vinokur, V. Lindemann criterion and vortex lattice phase transitions in type-II superconductors. *Phys. Rev. B* **69,** 024501 (2004).

[14] Menon, G. I. Phase behavior of type-II superconductors with quenched point pinning disorder: A phenomenological proposal. *Phys. Rev. B* **65,** 104527 (2002).

[15] Chudnovsky, E. M. Orientational and positional order in flux line lattices of Type-II superconductors. *Phys. Rev. B* **43,** 7831-7836 (1991).

[16] Paltiel, Y. *et al.* Instabilities and Disorder-Driven First-Order Transition of the Vortex Lattice. *Phys. Rev. Lett.* **85,** 3712-3715 (2000).

[17] Roy, S. B., Chaddah, P. & Chaudhary, S. Peak effect in CeRu2: History dependence and supercooling. *Phys. Rev. B* **62,** 9191-9199 (2000).

[18] Zeldov, E. *et al.* Thermodynamic observation of first-order vortex-lattice melting transition in $Bi_2Sr_2CaCu_2O_8$. *Nature* **375,** 373-376 (1995).

[19] Soibel, A. *et al.* Imaging the vortex-lattice melting process in the presence of disorder. *Nature* **406,** 282-287 (2000).

[20] Sarkar, S. *et al.* Multiple magnetization peaks in weakly pinned $Ca_3Rh_4Sn_{13}$ and $YBa_2Cu_3O_{7-\delta}$. *Phys. Rev. B* **64,** 144510 (2001).

[21] Banerjee, S. S. *et al.* Peak effect, plateau effect, and fishtail anomaly: The reentrant amorphization of vortex matter in 2H−$NbSe_2$. *Phys. Rev. B* **62,** 11838-11845 (2000).

[22] Safar, H. *et al.* Experimental evidence for a multicritical point in the magnetic phase diagram for the mixed state of clean untwinned $YBa_2Cu_3O_7$. *Phys. Rev. Lett.* **70,** 3800-3803 (1993).

[23] Mohan, S. *et al.* Large Low-Frequency Fluctuations in the Velocity of a Driven Vortex Lattice in a Single Crystal of 2H-NbSe2 Superconductor. *Phys. Rev. Lett.* **103,** 167001 (2009).

[24] Ghosh, K *et al.* Reentrant Peak Effect and Melting of a Flux Line Lattice in 2H-$NbSe_2$. *Phys. Rev. Lett.* **76,** 4600-4603 (1996).

[25] Banerjee, S. S. *et al.* Metastability and switching in the vortex state of 2H-$NbSe_2$. *Appl. Phys. Lett.* **74,** 126-128 (1999).

[26] Ravikumar, G. *et al.* Stable and metastable vortex states and the first-order transition across the peak-effect region in weakly pinned 2H-$NbSe_2$. *Phys. Rev. B* **63,** 024505 (2000).

[27] Pastoriza, H., Goffman, M. F., Arribére, A., & de la Cruz, F. First Order Phase Transition at the Irreversibility Line of $Bi_2Sr_2CaCu_2O_{8+\delta}$. *Phys. Rev. Lett.* **72,** 2951-2954 (1994).

[28] Larkin, A. A., & Ovchinnikov, Y. Pinning in type II superconductors. *J. Low Temp. Phys.* **34,** 409-428 (1979).

[29] D'Anna, G. *et al.* Flux-line response in 2H-NbSe2 investigated by means of the vibrating superconductor method. *Physica C* **218**, 238-244 (1993).

[30] Petrović, A. P. *et al.* Real-Space Vortex Glass Imaging and the Vortex Phase Diagram of $SnMo_6S_8$. *Phys. Rev. Lett.* **103,** 257001 (2009).

[31] Troyanovski, A. M., van Hecke, M., Saha, N., Aarts, J. & Kes, P. H. STM Imaging of Flux Line Arrangements in the Peak Effect Regime. *Phys. Rev. Lett.* **89,** 147006 (2002).

[32] Hecher, J., Zehetmayer, M. & Weber, H. W. How the macroscopic current correlates with the microscopic flux-line distribution in a type-II superconductor: an experimental study. *Supercond. Sci. Technol.* **27,** 075004 (2014).

[33] Suderow, H., Guillamón, I., Rodrigo, J. G. & Vieira, S. Imaging superconducting vortex cores and lattices with a scanning tunneling microscope. *Supercond. Sci. Technol.* **27,** 063001 (2014).

[34] Guillamón, I. *et al.* Enhancement of long-range correlations in a 2D vortex lattice by an incommensurate 1D disorder potential. *Nat. Phys.* **10**, 851-856 (2014).

[35] Iavarone, M. *et al.* Effect of magnetic impurities on the vortex lattice properties in $NbSe_2$ single crystals. *Phys. Rev. B* **78,** 174518 (2008).

[36] Marchevsky, M., M. Higgins, M. & Bhattacharya, S. Two coexisting vortex phases in the peak effect regime in a superconductor. *Nature* **409,** 591-594 (2001).

[37] Gammel, P. L. *et al.* Structure and correlations of the flux line lattice in crystalline Nb through peak effect. *Phys. Rev. Lett.* **80,** 833-836 (1998).

[38] U. Yaron et al., Structural evidence for a two-step process in the depinning of the superconducting flux-line lattice. *Nature* **376,** 753-755 (1995).

[39] Kosterlitz, J. M. & Thouless, D. J. Early work on Defect Driven Phase Transitions, *in 40 years of Berezinskii-Kosterlitz-Thouless Theory, ed. Jorge V Jose (World Scientific, 2013).*

[40] Giamarchi, T. & Le Doussal, P. Phase diagram of flux lattices with disorder. *Phys. Rev. B* **55,** 6577 (1997).

[41] Murray, C. A., Gammel, P. L., Bishop, D. J., Mitzi, D. B. & Kapitulnik, A. Observation of a hexatic vortex glass in flux lattices of the high-Tc superconductor $Bi_{2.1}Sr_{1.9}Ca_{0.9}Cu_2O_{8+\delta}$. *Phys. Rev. Lett.* **64,** 2312-2315 (1990).

[42] Kamlapure, A. *et al.* A 350 mK, 9 T scanning tunneling microscope for the study of superconducting thin films on insulating substrates and single crystals. *Rev. Sci. Instrum.* **84**, 123905 (2013).

[43] Horcas, I. *et al.* WSXM: A software for scanning probe microscopy and a tool for nanotechnology. *Rev. Sci. Instrum.* **78**, 013705 (2007).

**Figure Legends**

**Figure 1|** (a) Magnetic field *(H)* dependence of the real part of linear ac susceptibility ($\chi'$) (normalised to its value in zero field) at 350 mK for the VL prepared using different thermomagnetic cycling. The red line is $\chi'$-$H$ when the magnetic field is slowly ramped up after cooling the sample in zero field (ZFC state). The black line is $\chi'$-$H$ when the magnetic field is ramped down from a value higher than $H_{c2}$. The square symbols stand for the $\chi'$ for the FC states obtained by cooling the sample from $T > T_c$ in the corresponding field; the dashed line shows the locus of these FC states created at different $H$. The thin lines starting from the square symbols show the evolution of $\chi'$ when the magnetic field is ramped up or ramped down (ramped down segment shown only for 0.8 T), after preparing the VL in the FC state. We observe that the FC state is extremely unstable to any perturbation in magnetic field. $\chi'$ is normalised to the zero field value for the ZFC state. (b) Phase diagram showing the temperature evolution of $H_p^{on}$, $H_p$ and $H_{c2}$ as a function of temperature.

**Figure 2|** Representative STS conductance maps on $NbSe_2$ over 1 μm × 1 μm area at (a) 15 kOe and (b) 24 kOe. The conductance maps are recorded at d.c. bias voltage of 1.2 mV. The vortices are observed as local minima in the conductance map.

**Figure 3|** (a)-(f) STS conductance maps showing real space ZFC vortex lattice image at 350 mK along with their Fourier transforms. Delaunay triangulation of the VL are shown as solid lines joining the vortices and sites with 5-fold and 7-fold coordination are shown as red and white dots respectively. The disclinations (unpaired 5-fold or 7-fold coordination sites) observed at 26 and 30 kOe are highlighted with green and purple circles. While all images are acquired over 1 μm × 1μm area, images shown here have been zoomed to show around 600 vortices for clarity. The Fourier transforms correspond to the unfiltered images; the color scales are in arbitrary units. The images between 20-30 kOe were recorded in the same

magnetic field ramp, whereas the image at 15 kOe was recorded in a separate magnetic field ramp on the same cleaved surface. (g)-(h) VL images (400 nm × 400 nm) at 32 kOe and 34 kOe. At these fields the VL image becomes blurred due to the motion of vortices.

**Figure 4|** (a) Orientational correlation function, $G_6$ and (b) and positional correlation function, $G_K$ (averaged over the principal symmetry directions) as a function of $r/a_0$ for the ZFC state at various fields. $a_0$ is calculated by averaging over all the bonds after Delaunay triangulating the image. In panel (b), $G_K$ for each successive fields are separated by adding a multiple of one for clarity. (c) $G_6$ plotted in log-log scale for 20 kOe, 24 kOe and 25 kOe, along with fits (grey lines) of the power law decay of the envelope; $\eta_{or}$ are the exponents for power-law decay of $G_6$. (d) $G_K$ for 20 kOe, 24 kOe and 25 kOe plotted in semi-log scale along with the fits (grey lines) to the exponential decay of the envelope at short distance, with characteristic decay length, $\xi_p$. (e) $G_K$ for 24 and 25 kOe log-log scale along with fits (grey lines) to the power-law decay of the envelope at large distance; $\eta_p$ are the exponents for power-law decay of $G_K$. (f) $G_6$ plotted in semi-log scale for 26 kOe, 28 kOe, 30 kOe along with the fits (grey lines) to the exponential decay of the envelope with characteristic decay length, $\xi_{or}$. (g)-(h) Variation of the VL correlations lengths $\xi^{\parallel}$ and $\xi^{\perp}$ with magnetic field; below 15 kOe the correlation lengths reach the size of the image.

**Figure 5|** (a)-(c) VL images at 350 mK along with Delaunay triangulation for the ramp down branch at 25, 20 and 15 kOe. Sites with 5-fold and 7-fold coordination are shown as red and white dots respectively. At 25 and 20 kOe we observe dislocations in the VL. At 15 kOe the VL is topologically ordered. While all images are acquired over 1 μm × 1μm area, images shown here have been zoomed to show around 600 vortices for clarity. (d) $G_6$ and (e) $G_K$

calculated from the VL images at different field when the magnetic field is ramped down from $H > H_{c2}$ at 350 mK. The curves for each successive field are separated by adding a multiple of one for clarity. (f) $G_6$ at 25 kOe for ZFC and ramp down branch along with the fit (solid lines) to the power-law decay of the corresponding envelope; the exponents for the power-law fits, $\eta$, are shown in the legend.

**Figure 6|** Field Cooled VL images (upper panels) at 15 kOe, 20 kOe, 26 kOe, 28 kOe at 350 mK. Delaunay triangulation of the VL are shown as solid lines joining the vortices and sites with 5-fold and 7-fold coordination are shown as red and white dots respectively. The disclinations are highlighted with green and purple circles. The lower panels show the variation of $G_6$ for the FC and ZFC states at 26 kOe and 28 kOe along with the corresponding fits for exponential decay (solid lines); the decay lengths for the orientational order, $\xi_{or}$ are shown in the legend. Images have been zoomed to show around 600 vortices for clarity.

**Figure 7|** Annihilation of dislocations in a FC vortex lattice at 15 kOe with application of magnetic field pulse. (a) FC VL at 350 mK; the same VL after a applying a magnetic field pulse of (b) 0.3 kOe and (c) 0.9 kOe. Dislocations in the VL are shown as pairs of adjacent points with five-fold (red) and seven-fold (white) coordination. In (b) many of the dislocations are annihilated, whereas in (c) all dislocations are annihilated. Delaunay triangulation of the VL are shown as solid lines joining the vortices. Images have been zoomed to show around 600 vortices for clarity.

**Acknowledgements**

We thank Shobo Bhattacharya, Gautam Menon, Deepak Dhar and Rajdeep Sensarma for valuable discussions during the course of this work.

**Author Contributions**

SCG performed the STS measurements. SCG and GS analysed the STS data. RG and HS performed the bulk pinning measurements and analysed the data. PR conceptualised the problem, supervised the measurements and analysis and wrote the paper. PS, VB and AT carried out growth and characterisation of the samples. All authors discussed the results and commented on the manuscript.

**Competing financial interests:** The authors declare no competing financial intere

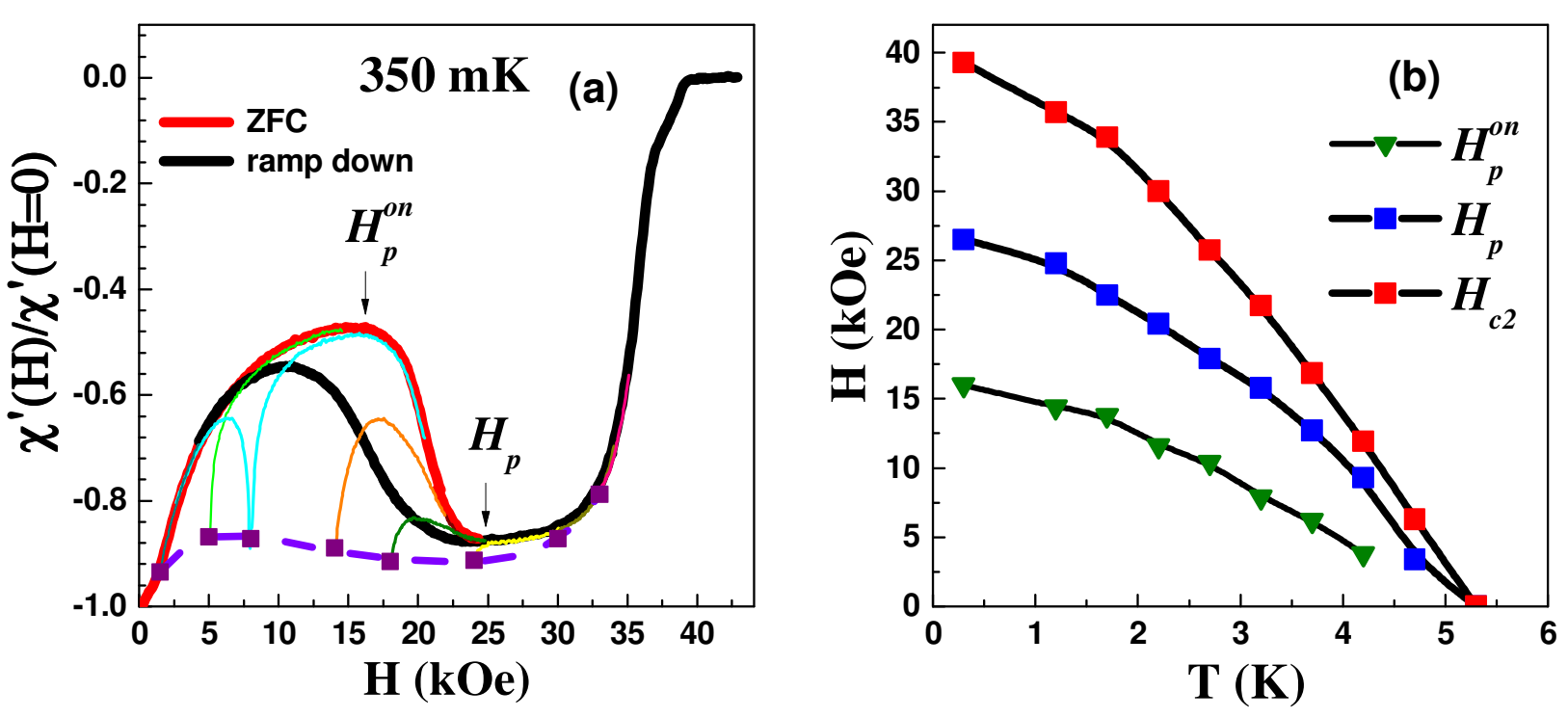

Fig. 1

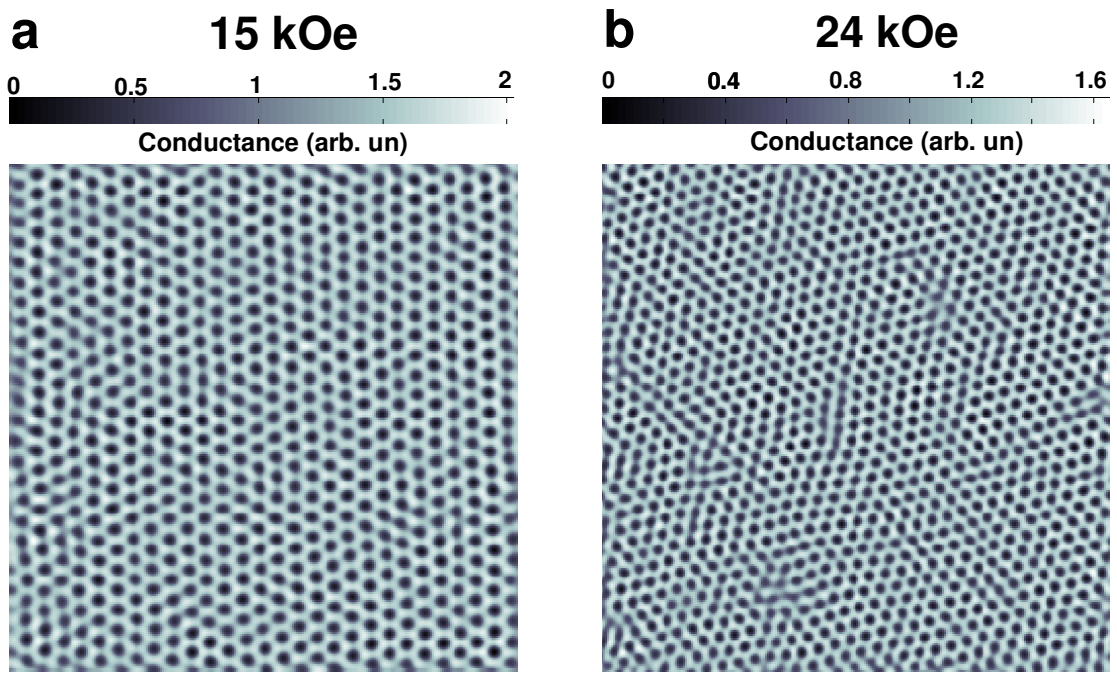

Fig. 2

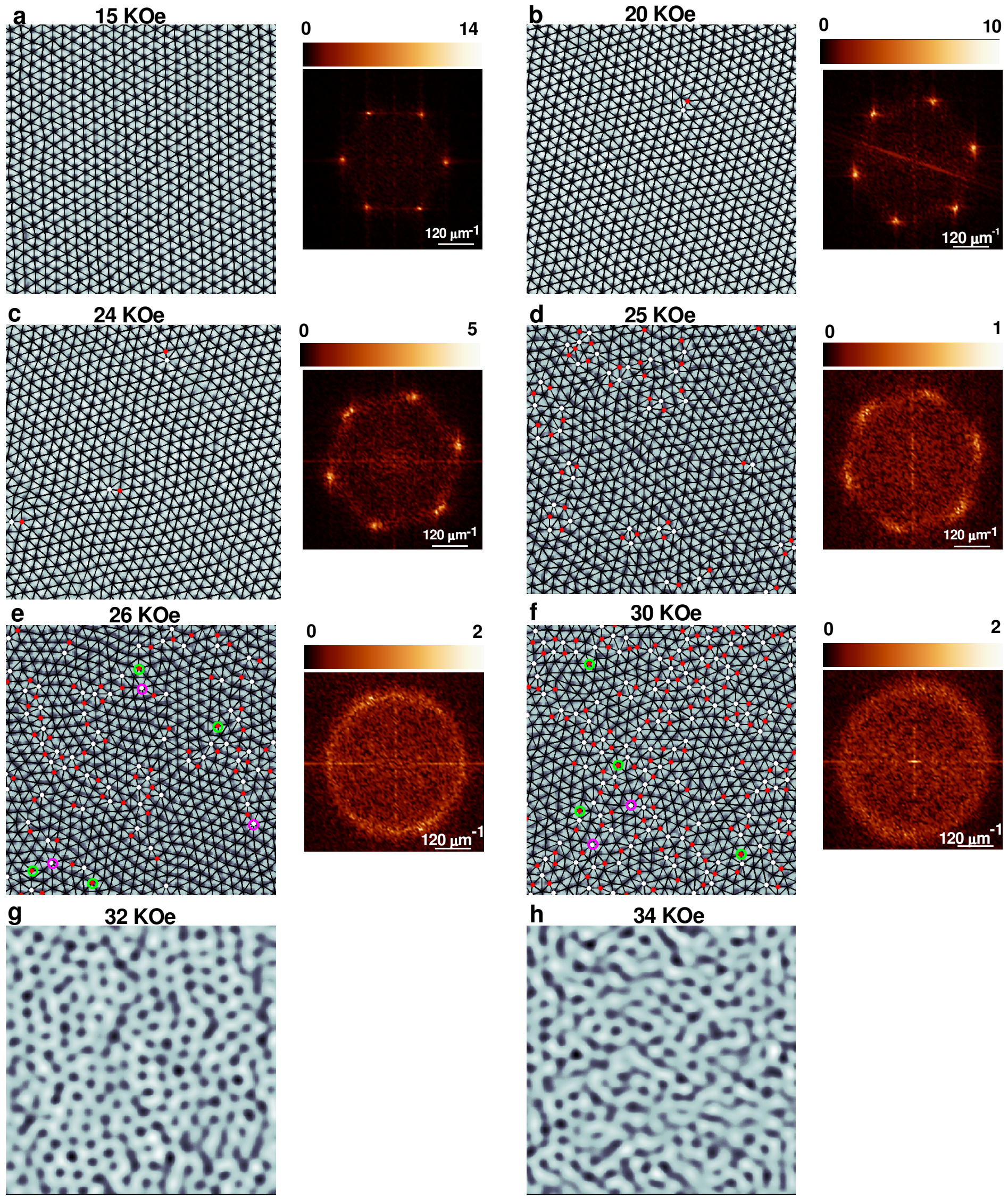

Fig. 3

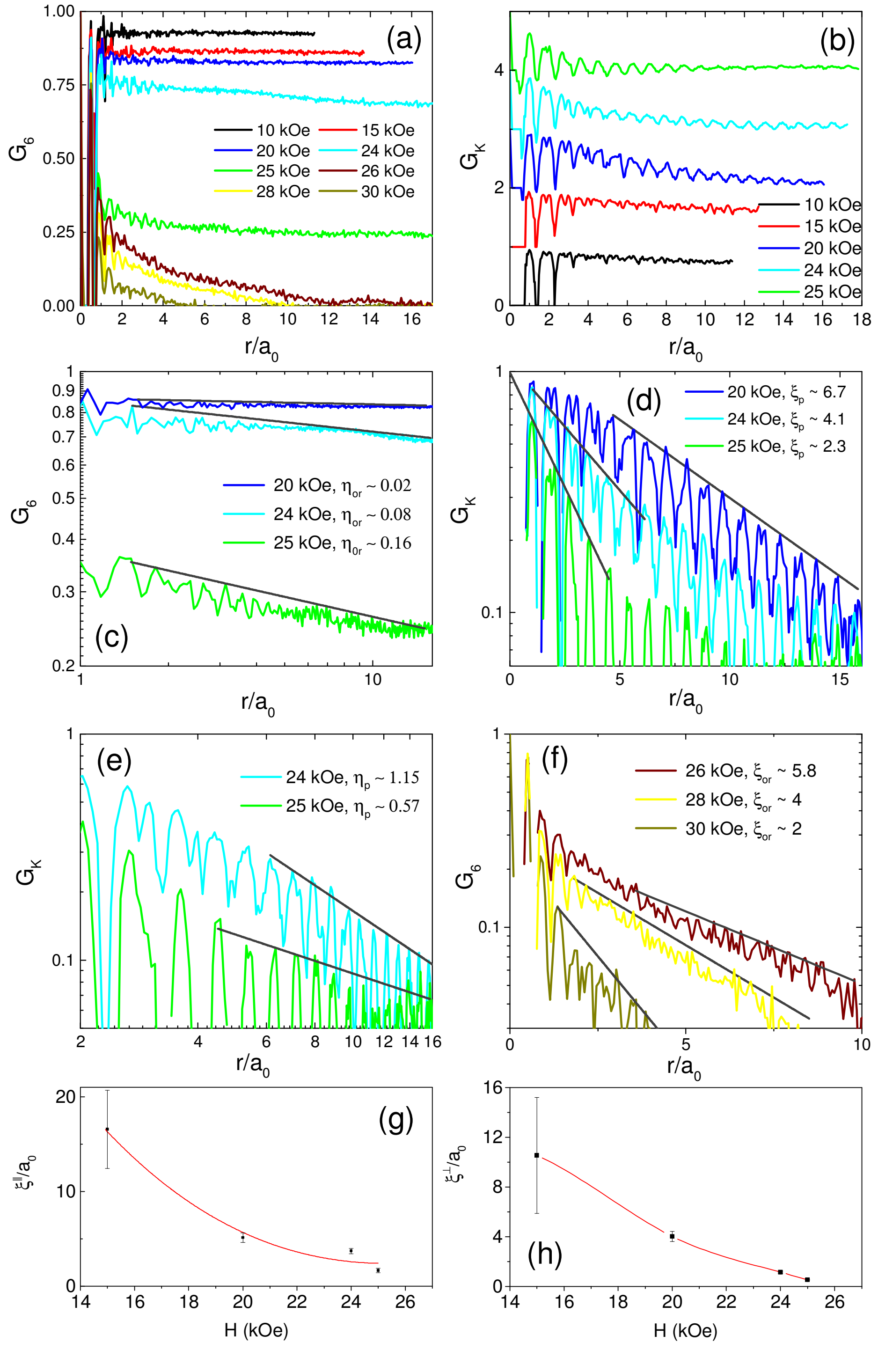

(a)
10 kOe
15 kOe
20 kOe
24 kOe
25 kOe
26 kOe
28 kOe
30 kOe
G6
r/a0
(b)
GK
(c)
20 kOe, ηor ~ 0.02
24 kOe, ηor ~ 0.08
25 kOe, η0r ~ 0.16
(d)
20 kOe, ξp ~ 6.7
24 kOe, ξp ~ 4.1
25 kOe, ξp ~ 2.3
(e)
24 kOe, ηp ~ 1.15
25 kOe, ηp ~ 0.57
(f)
26 kOe, ξor ~ 5.8
28 kOe, ξor ~ 4
30 kOe, ξor ~ 2
(g)
ξ∥/a0
H (kOe)
(h)
ξ⊥/a0

Fig. 4

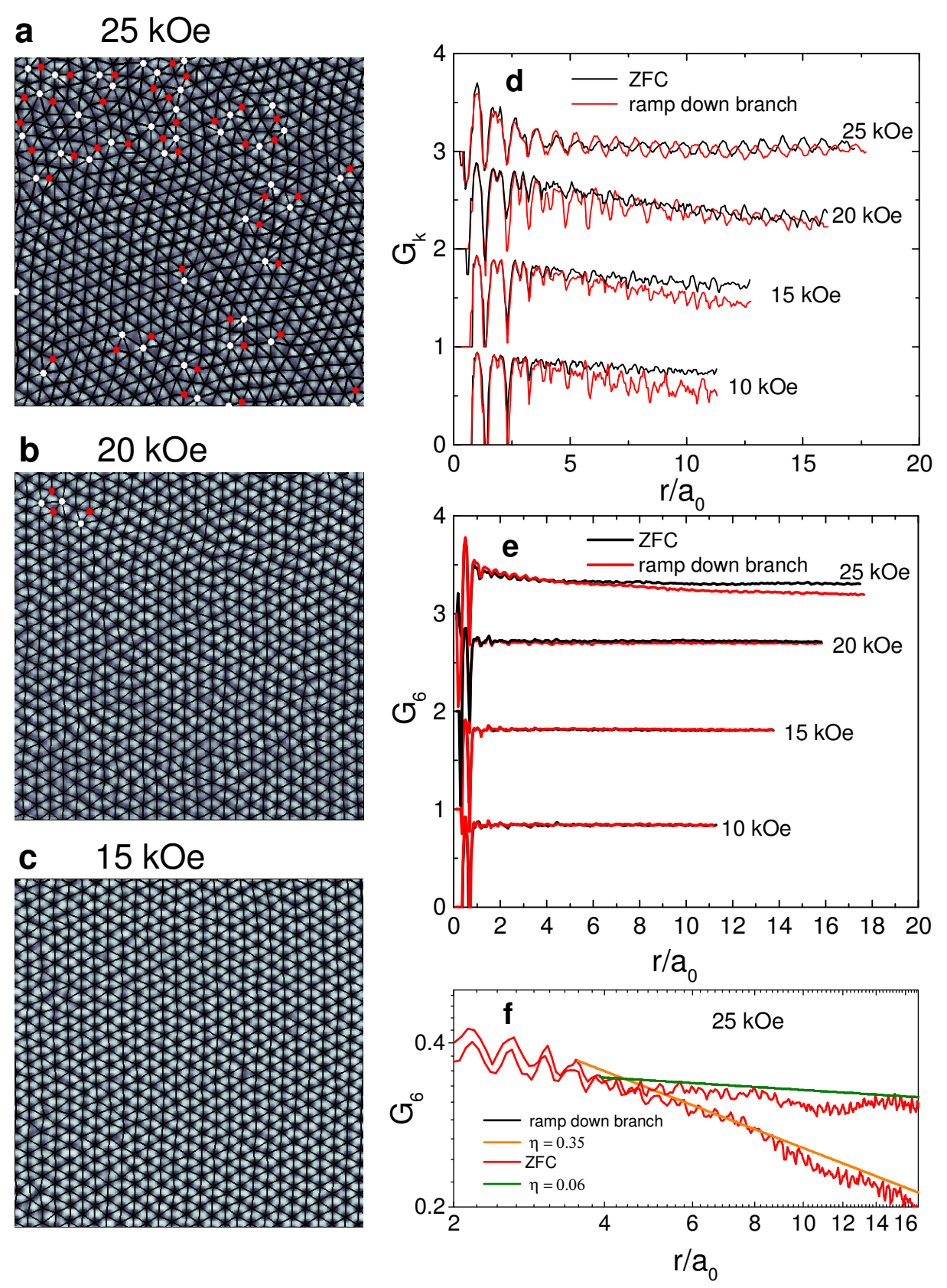

Fig. 5

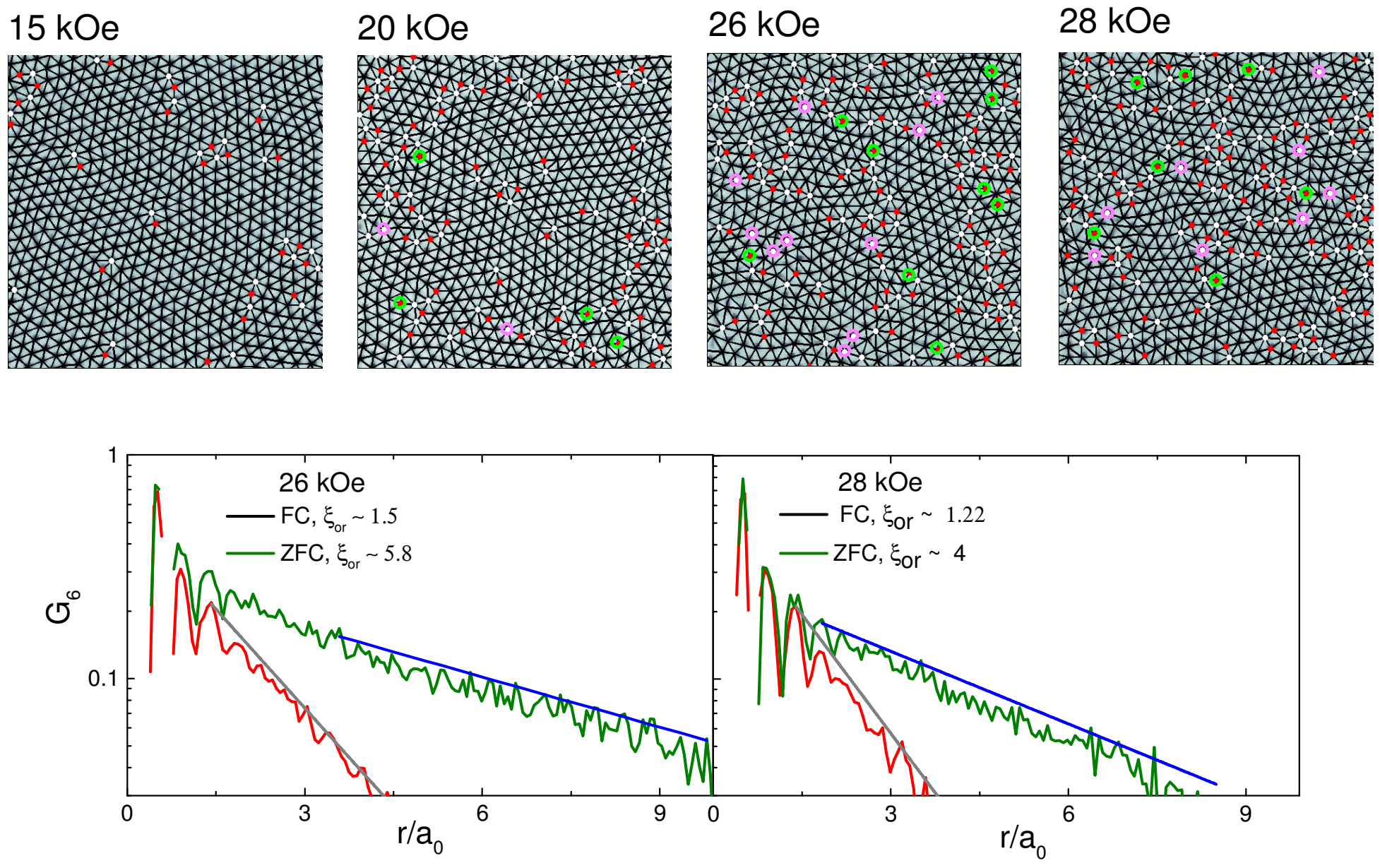
15 kOe
20 kOe
26 kOe
28 kOe
26 kOe
FC, $\xi_{or}$ ~ 1.5
ZFC, $\xi_{or}$ ~ 5.8
28 kOe
FC, $\xi_{or}$ ~ 1.22
ZFC, $\xi_{or}$ ~ 4
$G_6$
$r/a_0$

Fig. 6

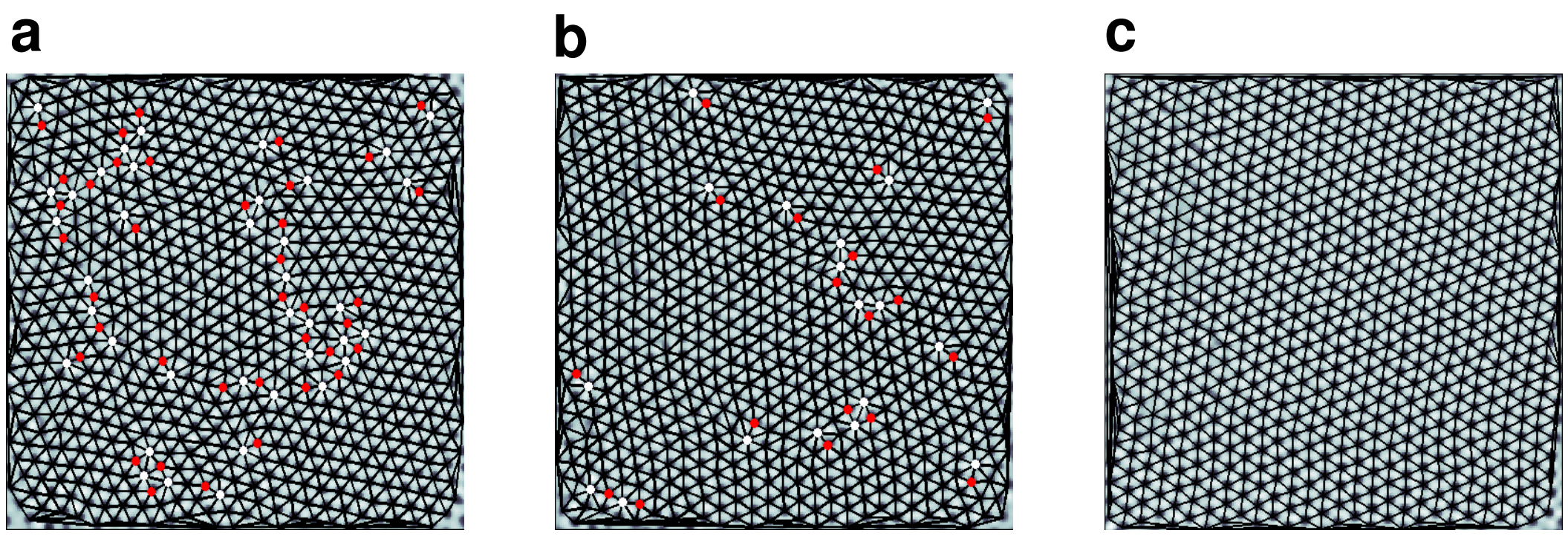

Fig. 7

**Two step disordering of the vortex lattice across the peak effect in a 3-dimensional type II superconductor $Co_{0.0075}NbSe_2$**

Somesh Chandra Ganguli[a], Harkirat Singh[a], Garima Saraswat[a], Rini Ganguly[a], Vivas Bagwe[a], Parasharam Shirage[b], Arumugam Thamizhavel[a] and Pratap Raychaudhuri[a,1]

[a] *Tata Institute of Fundamental Research, Homi Bhabha Road, Colaba, Mumbai 400005, India.*

[b] *Indian Institute of Technology Indore, IET-DAVV Campus, Khandwa Road, Indore 452017, India.*

## I. Temperature dependence of resistance of $Co_{0.0075}NbSe_2$ single crystal

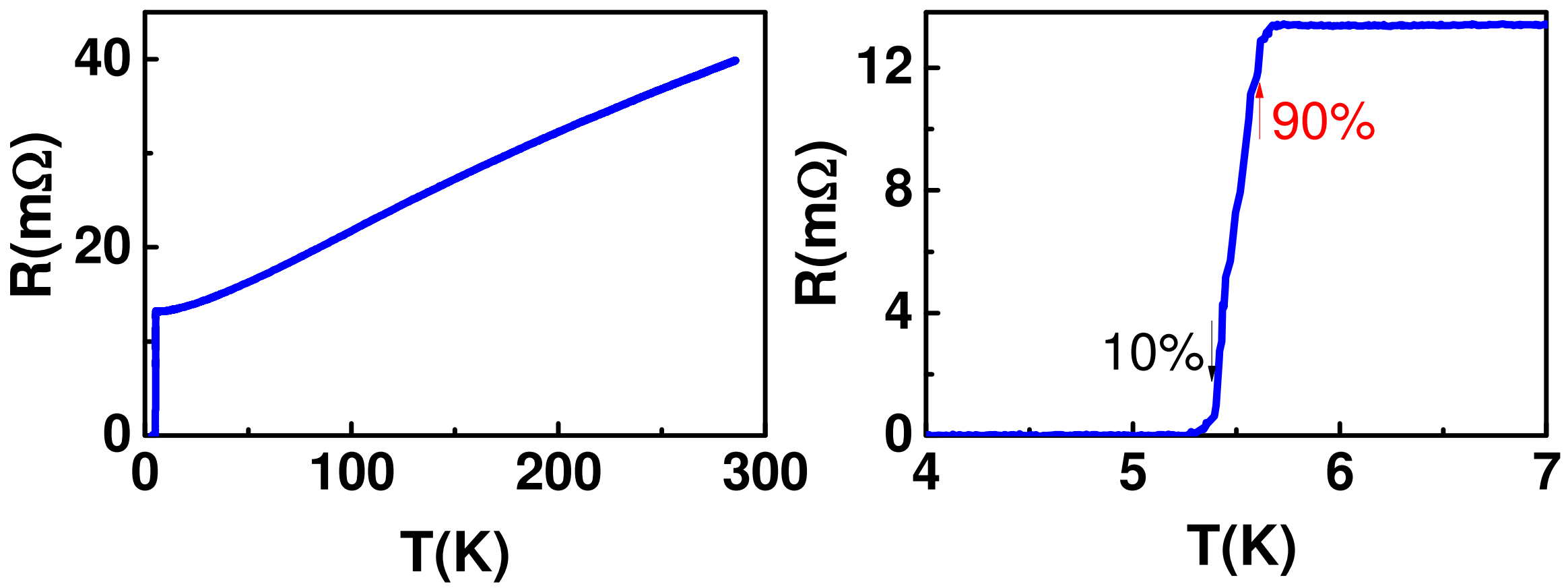

**Fig S1|** Resistance vs. temperature for the $Co_{0.0075}NbSe_2$ sample. The left panel shows the resistivity from 286 - 4 K. The right panel shows an expanded view close to the superconducting transition. The arrows mark the positions where the resistance is 90% and 10% of the normal state value respectively.

The $Co_{0.0075}NbSe_2$ single crystal was characterised using 4-probe resistivity measurements from 286 K to 4 K (Fig. S1). The superconducting transition temperature, defined as the temperature where the resistance goes below our measurable limit is 5.3 K. The transition width, defined as the difference between temperatures where the resistance is 90% and 10% of the normal state resistance respectively is ~ 200 mK. The same crystal was used for both a.c. susceptibility and STS measurements.

[1] E-mail: pratap@tifr.res.in

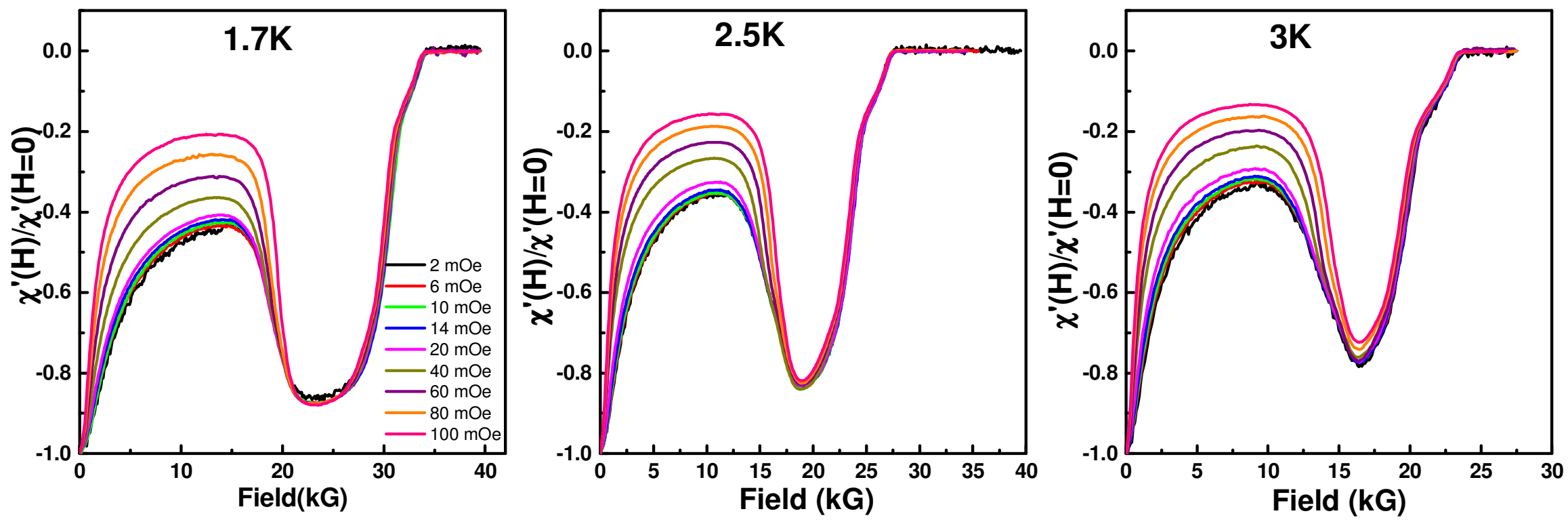

**Fig S2|** Isothermal χ'-H scans for different a.c. excitations. The susceptibility response shows a strong amplitude dependence above 14 mOe.

## II. a.c. susceptibility response as a function of excitation field

The a.c. susceptibility reported in Fig. 1 was performed with an a.c. excitation amplitude of 10 mOe at frequency 31 kHz. Since at large amplitudes, the a.c. drive can significantly modify the susceptibility response of the VL through large scale rearrangement of vortices, we performed several measurements with different a.c. excitation amplitudes to determine the range of a.c. field over which the χ' is independent of excitation field. We observe (Fig. S2) that below 3K, χ' shows significant dependence on the magnitude of the a.c. excitation only above 14 mOe.

## III. Calculation of the correlation lengths $\xi^{\|}$ and $\xi^{\perp}$ from VL images

For an infinite lattice the correlation lengths along and perpendicular to the reciprocal lattice vectors **K** can be obtained from the width of the first order Bragg peaks (BP) of the reciprocal lattice using the relations, $\xi^{\|} = 1/\Delta k_{\|}$ and $\xi^{\perp} = 1/\Delta k_{\perp}$, where $\Delta k_{\|}$ and $\Delta k_{\perp}$ are the width of the first order Bragg peaks parallel and perpendicular to **K**. This method has been used to determine the correlation lengths from the Bragg spots in neutron diffraction measurements. However, when using the Fourier transform of VL images obtained from STS measurements, additional precaution has to be taken to account for contributions arising from the finite size of images and any inaccuracy in position resulting from the finite pixel resolution.

To obtain $\xi^{\|}$ and $\xi^{\perp}$ from our data, at every field a binary lattice is first constructed using the position of each vortex (Fig. S3(a)) obtained the VL images as explained in the Methods section.

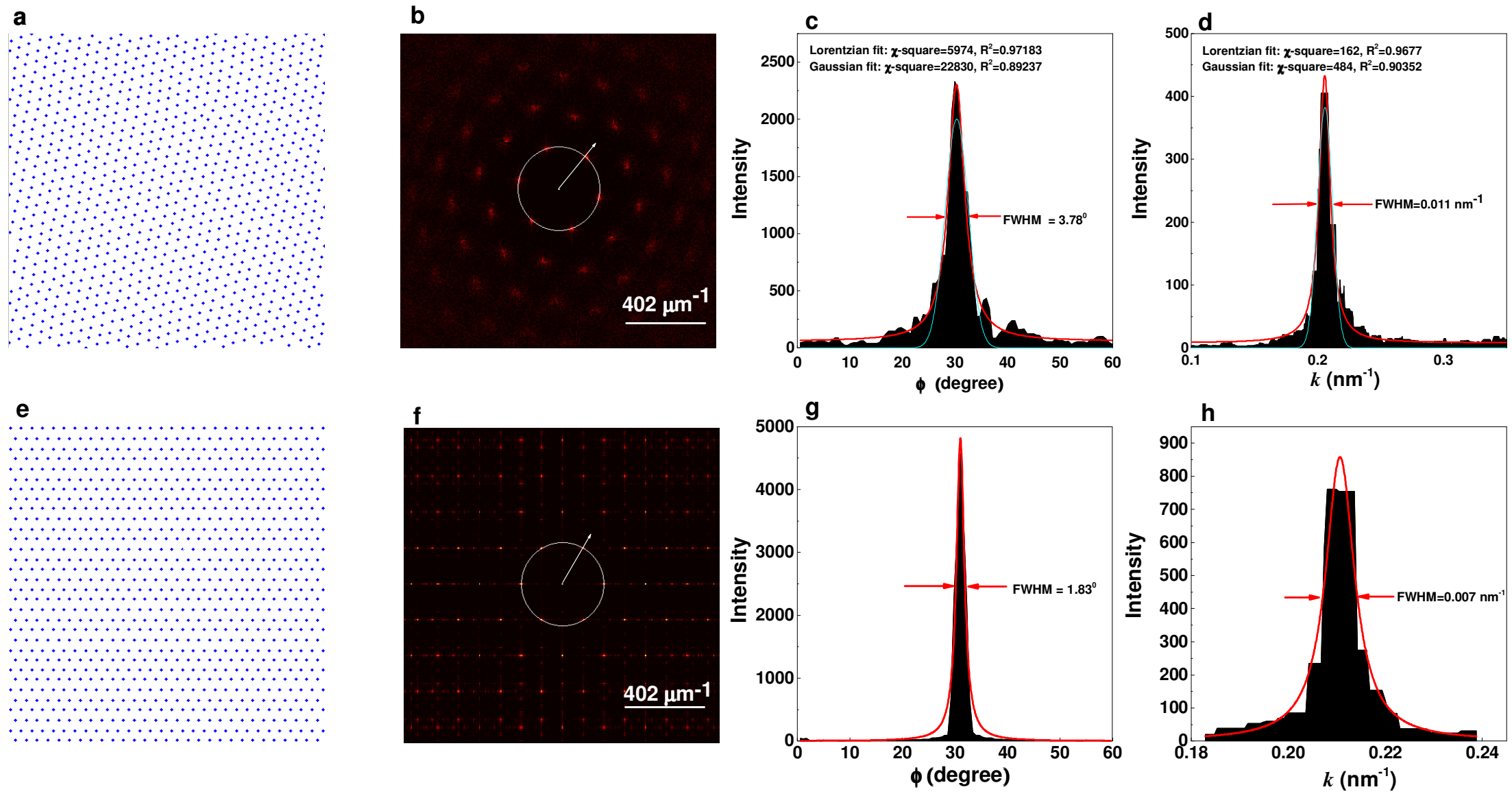

**Fig. S3|** (a) Binary image of the vortex positions constructed from the VL image at 20 kOe. (b) 2D Fourier transform of (a) showing one reciprocal lattice vector, **K**. (c) Azimuthal profile of the first order Bragg peak averaged over the six symmetric peak along the circle shown in (b); $\phi$ is the azimuthal angle with respect to an arbitrary axis roughly midway between two Bragg peaks. (d) Radial profile of the First order Bragg peak along the reciprocal lattice vector averaged over the six symmetric peaks; $k_{||}$ is the magnitude of wave-vector along the reciprocal lattice vector. The lines show the fit to the Lorentzian (red) and the Gaussian (indigo) functions; the corresponding $\chi$-square and $R^2$ values are shown in the panel. (e)-(h) Same as (a)-(d) for the ideal hexagonal lattice of same size with the same density of lattice points.

This removes image specific features and leaves only the precise position of each vortex within the uncertainty of pixel resolution. The 2D Fourier transform of this image reveals the first order Bragg peaks which appear as six symmetric bright spots as well as higher order peaks (Fig. S3(b)). (The position of the first order peaks with respect to the origin correspond to the reciprocal lattice vector, **K**.) We determine the peak profile along the azimuthal direction by taking the cut along a circle going through the center of the six first order Bragg spots. Similarly, the profile along the radial direction is determined by taking a line cut along the direction of the reciprocal lattice vector. We tried to fit peak profile averaged over the six symmetric BPs with both Lorentzian and Gaussian functions. The goodness of fit can be estimated from the residual sum of squares ($\chi$-square) and the coefficient of determination[1] (adjusted $R^2$) which should be 1 when the fit is perfect. We observe that the Lorentzian function gives a much better fit with smaller $\chi$-square and $R^2$ value close to unity. Therefore we calculate the corresponding peak width from the full width at half maxima of the best fit Lorentzian function (Fig. S3(c)-(d)).

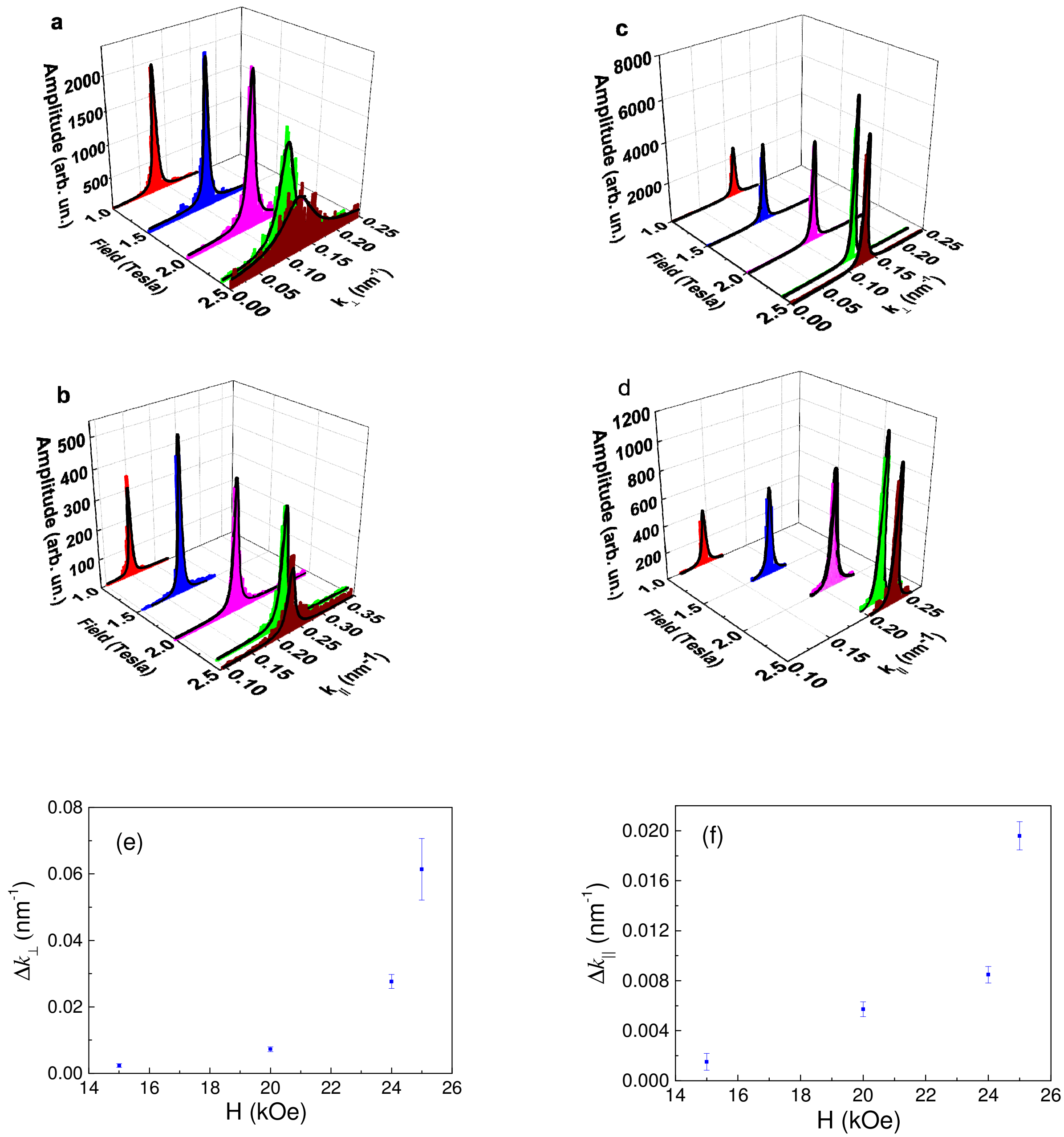

**Fig. S4|** (a)-(b) Profiles of the first order Bragg peaks (averaged over the six symmetric peaks) along the radial and azimuthal directions at different fields obtained from the VL images shown in Fig. 2 of the main paper. $k_{\perp} = \phi_0$ (in radian) × |**K**| is the distance in reciprocal space measured along the azimuthal direction. (c)-(d) Corresponding peak profiles for an ideal hexagonal lattice of the same size and same density of lattice points. The black lines are the fits to a Lorentzian function. (e)-(f) Variation of $\Delta k_{\perp}$ and $\Delta k_{||}$ with magnetic field.

Three factors contribute to the peak width determined in this process: (i) The intrinsic disorder in the lattice, (ii) the finite size of the image and (iii) the positional uncertainty arising from the finite pixel resolution of our images. For each field we construct the binary image of an ideal hexagonal lattice of the same size with the same density of lattice points, where the position of each lattice point is rounded off to same accuracy as the pixel resolution of our image (256 × 256). The radial and the azimuthal width of the first order Bragg spots for this ideal lattice is determined using the same procedure as before (see Fig. S3(e)-(h)). In principle the experimental peak width is a convolution of intrinsic and extrinsic factors, and to correct for the contributions arising from extrinsic factors one needs to follow an elaborate deconvolution procedure. However, when the peak can be fitted with a pure Lorentzian function, the situation is simpler and the peak widths arising from different contributions are additive. Since in our case we can fit the Bragg peaks with a pure Lorentzian function, we subtract the peak width of the ideal lattice from the peak width obtained from the actual image, to obtain $\Delta k_\perp$ and $\Delta k_\parallel$ arising from the lattice disorder alone.

Fig. S4 shows the evolution of azimuthal and radial peak profile averaged over the six first order Bragg spots as a function of magnetic field for the actual (Fig. S4(a)-(b)) and the ideal lattice (Fig. S4(c)-(d)). Fig. S4(e)-(f) show the magnetic field variation of the $\Delta k_\perp$ and $\Delta k_\parallel$ extracted from this data. We note that the evaluation of the correlations lengths from the relations, $\xi^{\parallel} = 1/\Delta k_\parallel$ and $\xi^{\perp} = 1/\Delta k_\perp$ is only valid up to the size of the image. Consequently, for our data $\xi^{\parallel}$ and $\xi^{\perp}$ are evaluated only for fields of 15 kOe and higher.

**IV. Filtering the STS conductance maps**

For better visual depiction, the conductance maps obtained from STS are digitally filtered to remove the noise and scan lines which arise from the raster motion of the tip. The filtering procedure is depicted in Fig. S5. Fig. S5(a) shows the raw conductance map obtained at 24 kOe. To filter the image we first obtain the 2D Fourier transform (FT) of the image Fig. S5(b). In addition to six bright spots corresponding to the Bragg peaks we observe a diffuse intensity at small ***k*** corresponding to the random noise and a horizontal line corresponding to the scan lines. We first remove the noise and scan line contribution from the FT by suppressing the intensity along the horizontal line and the diffuse intensity within a circle at small ***k*** (Fig. S5(c)). The filtered image shown in Fig. S5(d) is obtained by taking a reverse FT Fig. 5(c).

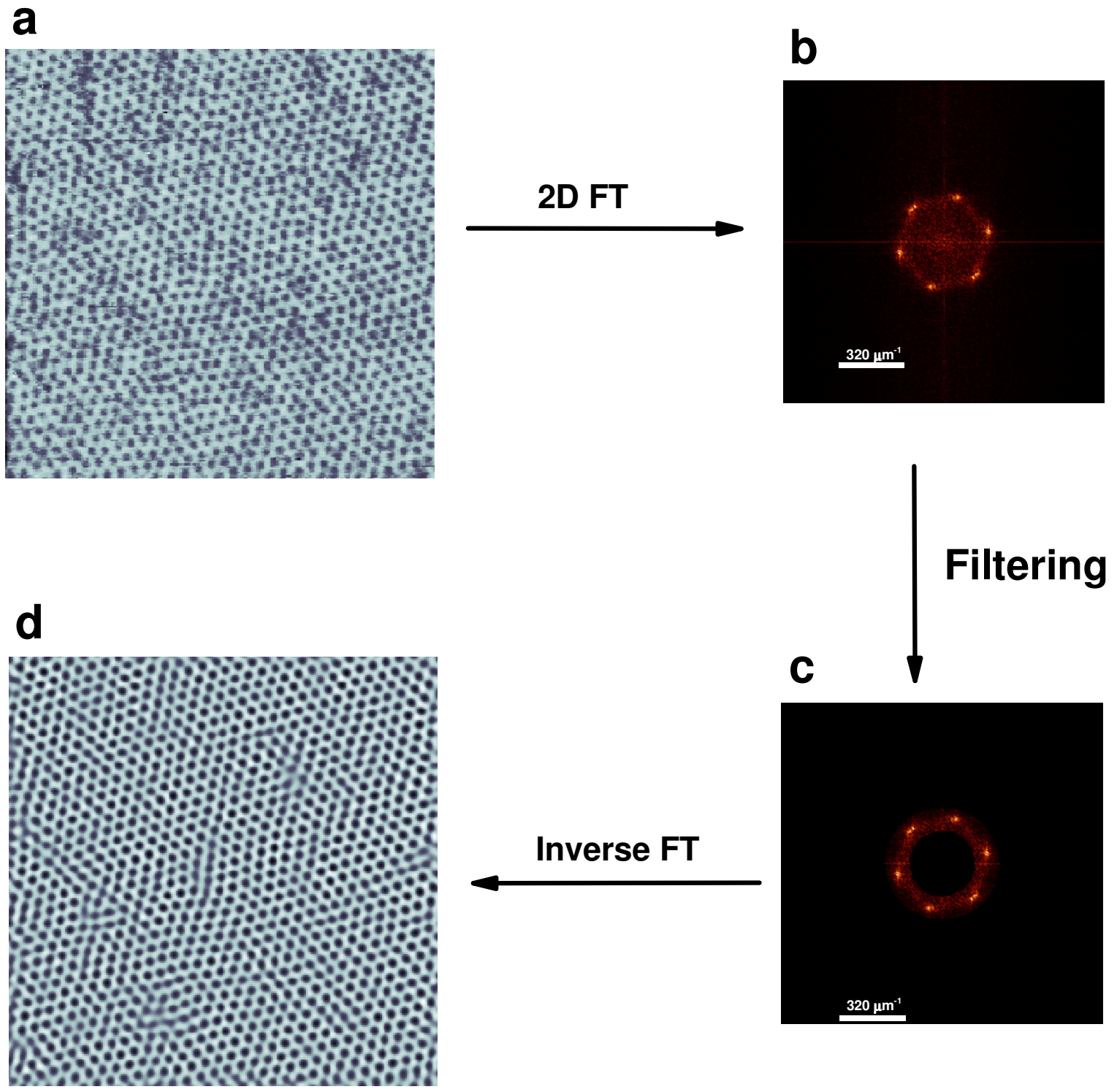

**Fig. S5|** Filtering the conductance map. (a) Raw conductance map recorded at 24 kOe, 350 mK over an area of 1 μm × 1 μm; (b) 2D FT of (a); (c) 2D FT after removing the contribution from random noise at small ***k*** and scan lines; (d) Filtered image obtained from the inverse FT of (c).

[1] Glantz, Stanton A.; Slinker, B. K. (1990). Primer of Applied Regression and Analysis of Variance. McGraw-Hill. ISBN 0-07-023407-8.